\documentclass[a4paper,11pt]{article}
\usepackage[T1]{fontenc}
\usepackage[utf8]{inputenc}
\usepackage{amsfonts}
\usepackage{amssymb}
\usepackage{amsmath}
\usepackage{lmodern}
\usepackage{graphicx}
\usepackage[usenames,dvipsnames]{color}
\usepackage{mathtools}
\usepackage{graphicx}

\usepackage{float}
\newtheorem{Theorem}{Theorem} 

\newtheorem{Definition}{Definition}
\newtheorem{Proposition}{Proposition}

\def\Real{{\mathbb R}}
\def\bcdot{{\mathbf \cdot}}
\renewcommand{\le}{\leqslant}
\renewcommand{\leq}{\leqslant}

\DeclareMathOperator{\supp}{supp}

\def\ellv{{L^2(\Sigma, \Real^N)}}

\title{Generalised hyperbolicity in spacetimes with Lipschitz regularity.}
\author{Yafet Sanchez Sanchez\footnote{E-mail:Y.SanchezSanchez@soton.ac.uk} \\ 
and James A. Vickers\footnote{E-mail:J.A.Vickers@soton.ac.uk} \\ Mathematical Sciences and STAG Research Centre, \\ University of Southampton,\\ Southampton,\\ SO17 1BJ } 

\begin{document}

\maketitle
\begin{abstract}

  In this paper we obtain general conditions under which the wave
  equation is well-posed in spacetimes with metrics of Lipschitz
  regularity. In particular, the results can be applied to spacetimes
  where there is a loss of regularity on a hypersurface such as
  shell-crossing singularities, thin shells of matter and surface
  layers. This provides a framework for regarding gravitational
  singularities not as obstructions to the world lines of
  point-particles, but rather as obstruction to the dynamics of test
  fields.
\end{abstract}

\begin{section}{Introduction}

  An important requirement of any classical physical theory is that
  given suitable initial data one can determine the evolution of the
  system. Within the theory of general relativity the concept of
  global hyperbolicity \cite{leray} therefore plays a key
  role. Mathematically, a spacetime region ${\mathcal N}$ is said to
  globally hyperbolic if the causality condition is satisfied, and for
  any two points $p, q \in {\mathcal N}$ the causal diamond $J^+(p)
  \cap J^-(q)$ is compact and contained in ${\mathcal N}$
  \cite{bernal}. However from the physical point of view the important
  property of smooth globally hyperbolic spacetimes is that the evolution of
  the Einstein equations is well defined (see e.g. \cite{hawking} for
  details).

  One way of interpreting the compactness of $J^+(p) \cap J^-(q)$ is
  that this set ``does not contain any points on the edge of
  spacetime, i.e. at infinity or at a singularity''
  \cite[\S6.6]{hawking}. In this context the regularity usually
  considered in the region ${\mathcal N}$ is that the metric must be
  $C^{1,1}$(also denoted $C^{2-})$\footnote{A function $f$ on an open set $\cal{U}$
    of $\mathbb{R}^{n}$ is said to be Lipschitz or $C^{0,1}$
    if there is some constant $K$ such that for each pair of points
    $p,q\in {\cal{U}}$, $\arrowvert f(p)-f(q)\arrowvert\le K\arrowvert
    p-q\arrowvert$, where $\arrowvert p\arrowvert$ denotes the usual
    Euclidean distance. We denote by $C^{k,1}$ those
    functions where the $k$th derivative is a Lipschitz
    function.  }.  This criteria is based on the existence and
  uniqueness of geodesics and the characterisation of 
  gravitational singularities in terms of geodesic
  incompleteness. Moreover this regularity is the threshold where
  causal theory for smooth metrics and rough metrics agree
  \cite{singc11,cm}.
  
   However, in many realistic astrophysical situations
  one would like to solve Einstein's Field Equations with lower
  regularity. For example, when one models a jump in density at the
  boundary of a star. This led Clarke \cite{generalized} to
  define a notion of generalised hyperbolicity directly in terms of
  the local existence and uniqueness of the wave equation.

  Clarke's definition of generalised hyperbolicity was motivated by
  two things. Firstly there are a number of spacetimes in which points
  are removed due to the presence of weak singularities and are
  therefore not globally hyperbolic but are still of physical
  interest. These include spacetimes with thin shells of matter
  \cite{thin} , impulsive gravitational waves \cite{penrose} and
  shell-crossing singularities \cite{Clarke&odonnell}. The second
  motivation was that of using test fields (given by solutions of the
  wave equation) rather than test particles (given by solutions of the
  geodesic equation) to probe the structure of singularities. Some
  work to examine the physical effect of gravitational singularities
  has been done by using a $3$-parameter family of test particles
  (see for example \cite{part}), however an advantage of using test
  fields is that the behaviour of the naturally defined
  energy-momentum tensor of the field gives a direct measurement of
  the physical effect of the singularity which is not so easy when
  considering families of test particles.

  There have been previous approaches which study the nature of the
  singularities using test fields (see e.g. \cite{wald,Marolf,kay, IH}
  ). In all these approaches one considers self-adjoint extensions of
  the (spatial) Laplace-Beltrami operator and applies appropriate
  boundary conditions at the singularity. All this work has focused on
  the case of static spacetimes and furthermore most of these
  self-adjoint extensions are in $L^2$ (except \cite{IH} who consider
  finite energy solutions).

  The approach in this paper is to consider singularities as interior
  points of a spacetime with low regularity rather than regarding the
  singularity as a boundary and applying boundary conditions. A
  natural condition in this context is to require the solutions to lie
  in the Sobolev space $H^1_{loc}$ which ensures that the energy
  momentum tensor is well-defined as a distribution. In this paper we
  will therefore look at conditions on the spacetime and initial data
  which give solutions with this regularity.
  
  In particular, we will look at the co-dimension one case which is
  relevant for example to junction conditions with jumps in the
  density, impulsive gravitational waves and brane-world
  cosmologies. In a previous paper we looked at the case of
  co-dimension two singularities such as cosmic strings \cite{ys2}.

  In previous work Grant et al \cite{GMS09} showed the existence of
  generalised solutions to the wave equation on singular
  spacetimes. This involved replacing the singular metric by a
  1-parameter family of smooth metrics. The 1-parameter family of
  solutions of the corresponding wave equations then describes a
  generalised solution. Using this approach it was possible to
  demonstrate the existence of a generalised solution to the wave
  equation for a wide class of metrics which included metrics with
  components bounded almost everywhere. A downside of this approach is
  that it is not easy to relate the generalised solution to a
  classical weak solution of the equation. In our approach we will
  utilise the vanishing viscosity method (see e.g. Evans \cite{evans})
  to show the existence of a weak solution of the wave equation. This
  involves approximating the wave equation by a system of second order
  parabolic equations with a parameter $\epsilon$ corresponding to the
  viscosity. Like Grant et al we again obtain a 1-parameter family of
  solutions but we are able to utilise results from parabolic
  regularity theory to have better control over the solutions and
  their convergence. This enables us to show the 1-parameter family
  converges to a weak solution of the zero viscosity equation which
  corresponds to an $H^1$ solution of the wave equation. The basic
  method we use follows that of Evans \cite{evans} but the details
  differ and our results are also distinct from from his since we
  assume less regularity and as a result our solutions also exhibit
  less regularity.
  
  The plan of the paper is as follows. In \S2 we introduce the general
  setting for the problem and state the main theorems. We do not
  impose the ultrastatic condition used in \cite{VW} and write the
  wave equation as a first order system (see
  e.g. \cite{ringstrom}). This enable us to work with the $L^2$
  energy of the first order system which corresponds to an $H^1$
  energy of the second order system. In this setting we obtain
  existence of the solutions in time dependent scenarios without the
  need for any symmetries. \S3 contains the proofs of the theorems
  while \S4 gives examples of how these can be applied to various
  examples such as a discontinuity across a hypersurface, impulsive
  gravitational waves and brane-world cosmologies.

\end{section}

\begin{section}{The main results}

  The interest in co-dimension one singular submanifolds covers a
  variety of different interesting physical phenomena such as, surface
  layers \cite{thin}, impulsive gravitational waves \cite{penrose},
  and shell-crossing singularities \cite{cross} all of which fall
  outside the class of smooth globally hyperbolic spacetimes.
  Moreover, the mathematical analysis by Geroch and Traschen \cite{gt}
  of what now is called the class of Geroch-Traschen metrics and the
  subsequent analysis by Steinbauer and Vickers \cite{gerocht} using
  generalised functions gives co-dimension one singular submanifolds a
  robust mathematical background. In addition, recent proposals in
  semi-classical gravity and quantum gravity \cite{pams} suggest that
  the metric near the event horizon may present some loss of
  regularity. In this section we present techniques to prove local
  well-posedness of the wave equation in spacetimes with co-dimension
  one singularities subject to certain condition on the metric.

Clarke used the term {\it generalised hyperbolicity} to describe the
situation where one has a unique solution to the wave
equation. However we want to impose the slightly stronger condition of
well-posedness and in addition want to emphasise the role of the wave
equation. 
See \cite{SS} for a discussion of this and related terms.

\subsection{The general setting }

The geometric setting is a region
${\Sigma_{[0,T]}}=[0,T]\times\Sigma$, where $\Sigma$ is either a compact
closed $n$-dimensional manifold or an open, bounded set of
$\mathbb{R}^{n}$ with smooth boundary $\partial \Sigma$. In what
follows, for simplicity, we will only consider the former case. The
proof in the latter case follows by replacing $H^{1}(\Sigma)$ by
$H^{1}_{0}(\Sigma)$ and using the volume form given
by $dx^{n}$.  \\

Rather than considering the particular case of a spacetime with a
singular hypersurface, where the regularity of the metric drops below
$C^2$, we will consider a rough spacetime, where the spacetime metric $g_{ab}$ is only Lipschitz. We
will show that in this situation one has well-posedness (in a sense
made precise below) of the wave equation with weak solutions of
regularity $H^{1}(\Sigma_{[0,T]})$. In order to do this, we will
reformulate the wave equation as a first order symmetric hyperbolic
system and look for $L^2$ solutions of this system.

We therefore start by considering the first order initial value problem

\begin{eqnarray}\label{pde3}
 L{\bf{u}}= A^{0}\partial_{t}{\bf{u}}+A^{i}\partial_i{\bf{u}}+B{\bf{u}}&=&{\bf F}  \label{PDE1} \\
{\bf{u}}(0,\cdot)&=&{\bf{u}}_{0}(\cdot) \label{BC1}
\end{eqnarray}
In the above we have employed the Einstein summation convention where
$i$, $j$, $k$, etc. range over $1,2, \dots,n$. The unknown ${\bf u}$ and the
source term ${\bf F}$ are both $\Real^{N}$ valued functions on $\Sigma_{[0,T]}$, while
$A^0$, $A^i$ and $B$ are $N \times N$ matrix valued functions on
$\Sigma_{[0,T]}$. We will assume that $A^0$ and $A^i$ are symmetric and that
in addition $A^0$ is positive definite. 

In order for such a system to correspond to the wave equation given by a
Lipschitz metric, we will require that $A^0$ and $A^i$ have bounded
first derivatives and that $B$ is bounded, so that we require
\begin{equation*}
A^0 \in W^{1,\infty}(\Sigma_{[0,T]}, \Real^{N^2}), \quad
A^i \in W^{1,\infty}(\Sigma_{[0,T]}, \Real^{N^2}), \quad
B \in L^{\infty}(\Sigma_{[0,T]}, \Real^{N^2}).
\end{equation*}  

In the analysis below we will be working in spaces such as
$L^2(\Sigma)$. Rather than defining this in terms of a particular
coordinate system on $\Sigma$, we will introduce a background Riemannian
metric $h_{ij}$ on $\Sigma$ and let $\nu_h$ be the corresponding
volume form. We then define $L^2(\Sigma)$ to be the space of real
valued functions $g$ on $\Sigma$ such that $\int_\Sigma g^2 \nu_h <
\infty$ and we denote the associated inner product by
$(f,g)_{L^2(\Sigma)}=\int_\Sigma fg\nu_h$. Note that since $\Sigma$ is
compact $\nu_h$ is bounded from below and above. Furthermore if
$\Sigma$ is parallelizable (which for simplicity we will assume) there
is no loss of generality in taking $h_{ij}$ to be the flat metric. Note that in the three dimensional case, which is most relevant to
applications in general relativity, it is enough for $\Sigma$ to be
orientable for it to be parallelizable.

For the case of vector valued functions ${\bf v}$ on $\Sigma$, we define
$L^2(\Sigma, \Real^N)$ to be those ${\bf v}$ such that $\int_\Sigma
{\bf v} \bcdot {\bf v} \nu_h < \infty$. The corresponding inner
product on $L^2(\Sigma, \Real^N)$ is then given by
\begin{equation*}
({\bf v}, {\bf w})_{L^2(\Sigma, \Real^N)}=\int_\Sigma {\bf v}\bcdot {\bf w}\nu_h
\end{equation*}
Where there is no risk of confusion, we will write both the inner product in  
$L^2(\Sigma)$ and in $L^2(\Sigma, \Real^N)$ simply as $(\cdot,\cdot)_{L^2}$
The Sobolev spaces $H^1(\Sigma, \Real^N)$ etc. are defined in a
similar manner ( see\cite{evans} \S5.2, \cite{ringstrom}).

We also make use of the function space $L^2(\Sigma_{[0,T]})$ which is
defined by requiring that functions are square integrable on $[0,T]
\times \Sigma$ with respect to the volume form $dt \wedge
\nu_h$. However in the analysis below it is often convenient to think
of a function ${\bf v}(t,x)$ as a map from $[0,T]$ to a function ${\bf
  v}(t)(\cdot)$ of $x \in \Sigma$ given by ${\bf v}(t)(x)={\bf v}(t,x)$. 
For example $L^2(0, T; L^2(\Sigma, \Real^N))$ is the space of functions 
\begin{eqnarray} \label{nested}
{\bf v}:[0,T] & \to & L^2(\Sigma, \Real^N)\\
t & \mapsto & {\bf v}(t)
\end{eqnarray} 
such that ${\bf v}(t) \in L^2(\Sigma, \Real^N)$ and
\begin{equation}
\int_0^T ({\bf v},{\bf v})_{\ellv}dt < \infty
\end{equation}
When thinking of ${\bf v}$ in this way, we will denote the time
derivative by ${\bf \dot v}$.

We also define the following spaces which will be used in \S3.4.
\begin{equation*}
  \tilde{W}^{k,p}({\Sigma_{[0,T]}},\Real^N)\coloneqq \{{\bf{w}}\in 
{W}^{k,p}({\Sigma_{[0,T]}},\Real^N): \text{such that } {\bf{w}}(T, \cdot)=0 \}
\end{equation*}

In particular, we will be interested in the following case

\begin{equation*}
  \tilde{H}^{1}({\Sigma_{[0,T]}},\Real^N)\coloneqq \{{\bf{w}}\in {H}^{1}({\Sigma_{[0,T]}},\Real^N): \text{such that } {\bf{w}}(T, \cdot)=0 \}.
\end{equation*}

\subsection{Weak solutions and the main theorems}

We will be looking for weak solutions of the initial value problem
(\ref{pde3}). To motivate the definition we proceed as follows.  Given
a standard $C^1$ solution ${\bf u}$ of the initial value problem, we
  may first take the dot product of equation (\ref{pde3}) with a smooth
  $\Real^N$-valued function ${\bf v}$ with support in $[0,T) \times
  \Sigma$ and then integrate over $x$ and $t$ to obtain
\begin{equation}
   \int^{T}_{0}(L{\bf u}, {\bf v})_{\ellv}dt=\int^{T}_{0}({\bf F},{\bf v})_{\ellv}dt
\end{equation}
Integrating the left hand side by parts with respect to $x$ and $t$ we
obtain
\begin{equation}
   \int^{T}_{0}({\bf u}, L^{*}{\bf v})_{\ellv}dt 
- \left(A^0{\bf{u}}|_{t=0}, {\bf v}(0)\right)_{\ellv}
=\int^{T}_{0}({\bf F},{\bf v})_{\ellv}dt
\end{equation}
where $L^{*}$ is the formal adjoint of $L$ defined below and the second term on the 
right hand side (RHS) comes from the $t=0$ boundary term when we integrate by parts
with respect to $t$. This approach results in the following
definition:
\begin{Definition}\label{ddeeff} {(\bf Weak Solution)}
We say a function:
$${\bf u} \in L^{2}(0,T;L^{2}(\Sigma,\mathbb{R}^{N}))$$
is a local weak solution of the initial value problem (\ref{pde3})
provided that:
\noindent
For all $ {\bf{{v}}}\in C^{\infty}({\Sigma_{[0,T]}},{\mathbb{R}^{N}})$,
with $\supp( {\bf{{v}}})\subseteq [0,T)\times\Sigma$
  \begin{equation}
   \int^{T}_{0}({\bf{u}}, L^{*} {\bf{{v}}})_{\ellv}dt
=\int^{T}_{0}({\bf F}, {\bf{{v}}})_{\ellv}dt 
+\left(A^{0}(0){\bf{u}}_0, {\bf{{v}}}(0)\right)_{\ellv}.
  \end{equation}
  
\end{Definition}
This definition of a weak solution is the classical one used by
Friedrichs \cite{fried} but differs from the one used by Evans
\cite{evans}, who does not integrate with respect to $t$. Note also
that the formal adjoint is defined with respect to $\nu_h$. So in the
case where we use a general Riemannian metric, there are additional
terms involving the derivatives of $\nu_h$ in the expression for
$L^{*}$ compared to the flat case. The explicit expression is:
\begin{equation} \label{adjoint}
  L^{*} {\bf{{v}}}\coloneqq -\partial_t(A^{0} {\bf{{v}}})-\partial_i(A^{i} {\bf{{v}}})+B^{T} {\bf{{v}}}-\tilde{\Gamma}^{i}_{li}A^{l}{\bf{{v}}}
\end{equation}
where $\tilde{\Gamma}^i_{jk}$ are the connection coefficients of the
smooth Riemannian metric $h_{ij}$.

In order to prove uniqueness and well-posedness of the initial value
problem, we will need to control the $L^2$ size of the solution. This
motivates the following definition.

\begin{Definition} \label{regular} {(\bf Regular Weak Solution}) We
  say a weak solution \\${\bf u} \in
  L^{2}(0,T;L^{2}(\Sigma,\mathbb{R}^{N}))$ is \textbf{\textsl{regular}} if
  ${\bf{u}}$ satisfies the energy estimate
\begin{equation}\label{eeE}
        ||{\bf{u}}||^{2}_{L^{2}(0,T;L^{2}(\Sigma,\mathbb{R}^{N}))}
\le C\left( ||{\bf{u}}_{0}||_{L^{2}(\Sigma,\mathbb{R}^{N})}
+\int^{T}_{0}||{\bf F}(t,\cdot )||^{2}_{L^{2}(\Sigma,\mathbb{R}^{N})} dt \right)    
\end{equation}
\end{Definition}

We may now state our main result concerning solutions of
low-regularity symmetric hyperbolic systems. 

\begin{Theorem}\label{tfo1}
Given the linear symmetric hyperbolic system:
\begin{eqnarray}
  L{\bf v}=A^{0}\partial_{t}{\bf{u}}+A^{i}\partial_i{\bf{u}}+B{\bf{u}}&=&{\bf F}\\
  {\bf{u}}(0,\cdot)&=&{\bf{u}}_{0}(\cdot)
\end{eqnarray}
where $A^0, A^i, B$ and ${\bf F}$ are as above, and the initial data
${\bf{u}}_{0}$ is in $L^{2}(\Sigma,\mathbb{R}^{N})$. Then
there exists a unique regular weak solution ${\bf{u}}\in
L^{2}(0,T;L^{2}(\Sigma,\mathbb{R}^{N}))$. Furthermore this solution is
stable in the sense that the solution depends continuously on the
norm of the initial data in $L^{2}(\Sigma,\mathbb{R}^{N})$ and the norm of the source function in $L^{2}(\Sigma_{[0,T]},\mathbb{R}^{N})$.
\end{Theorem}

We may now use the above result to establish the following theorem for
the wave equation.

\begin{Theorem}\label{tfo}
  Let $g_{ab},g^{ab}$ be in $C^{0,1} $,  and $f$ in
  $L^{2}(\Sigma_{[0,T]})$. Given initial data $(u_{0}, u_{1}) \in 
  H^{1}(\Sigma) \times L^{2}(\Sigma)$ then the system
  \begin{eqnarray}\label{lpde}
 \square_{g}u+m^{2}u&=&f\\
  u(0,\cdot)&=&u_{0}\\
  \partial_{t}u(0,\cdot)&=&u_{1}
 \end{eqnarray} 
 has a unique stable solution $u\in H^{1}({\Sigma_{[0,T]}})$. Moreover,
 the corresponding energy-momentum tensor $T_{ab}[u]$ is in
 $L^{1}_{loc}(\Sigma_{[0,T]})$.
\end{Theorem}
Note that the above result is similar to the one obtained for the
homogeneous wave equation in \cite{nicolas}. However, we have extended
the
 results to the more general case that includes mixed space and
time derivatives,included a source function and provided a different
proof.
 
 We would like to mention that one can not go further
without paying a price. 
 As shown by Colombini et al. \cite{sur}
there are examples of wave equations with coefficients depending only
on time of Holder regularity \footnote{ A function $f$ on an open set
  $\cal{U}$
 of $\mathbb{R}^{n}$ is said to be Holder or
  $C^{0,\alpha}$
 if there is some non negative constant $K$ such
  that for each pair of points
 $p,q\in {\cal{U}}$, $\arrowvert
  f(p)-f(q)\arrowvert\le K\arrowvert
 p-q\arrowvert^{\alpha}$, where
  $\arrowvert p\arrowvert$ denotes the usual
 Euclidean distance.}
$C^{0,\alpha}$ with exponent $\alpha<1$
 with no distributional
solution. However, they proved well-posedness by moving from Sobolev
spaces to Gevrey spaces. These results have been further extended to
coefficients that depend smoothly in the space variable but are
log-Lipschitz (LL) \footnote{A function $f$ on an open set $\cal{U}$ of
  $\mathbb{R}^{n}$ is said to be log-Lipschitz if there is some non
  negative constant $K$ such that for each pair of points
 $p,q\in
  {\cal{U}}$, $\arrowvert f(p)-f(q)\arrowvert\le K\arrowvert
  p-q\arrowvert\arrowvert
 \ln |p-q|\arrowvert$, where $\arrowvert
  p\arrowvert$ denotes the usual
 Euclidean distance.} regular in
time \cite{lerner}. In both \cite{sur} and \cite{lerner} the structure
of the second order part of the operator considered has the special form
 
    \begin{equation}
      \frac{\partial^{2}}{\partial t^{2}}-\displaystyle\sum_{j,k}\frac{\partial}{\partial x^{j}}\left(a_{j,k}(x,t) \frac{\partial}{\partial x^{k}}\right).
    \end{equation} 
    
   In \cite{men} mixed terms in time and space were allowed and the regularity of the coefficients was LL in space and time. However, the local well posedness results obtained are in Sobolev spaces $H^{s}$ with $|s|<1$ and therefore the energy momentum tensor of the solutions is not integrable. Finally, one can explore wave-type equations with very rough coefficients such as in \cite{rud} where one is lead to the necessity of weakening the notion of a solution to the Cauchy problem and enlarging the allowed class of solutions.

\section{Proof of the main theorem}

\subsection{Outline of the proof} 

The proof of Theorem \ref{tfo} uses the {\it vanishing viscosity
  method} described in $\S 7.3$ of Evans \cite{evans}. Note however
that Evans assumes that the $A^0$ and $A^i$ have greater regularity
than we do and as a result is able to obtain a solution with greater
regularity. This explains why our definition of a weak solution has to
differ from his. However the essence of the proof is essentially the
same. It consists of the following steps:

\begin{enumerate}
\item First, we approximate the problem (\ref{PDE1}) by the system of
  parabolic initial value-problem on ${\Sigma_{[0,T]}}$ given by
\begin{eqnarray}\label{PDE2}
  \partial_t{\bf{u}}^{\epsilon}-\epsilon\Delta_{h} {\bf{u}}^{\epsilon}+
(A^{0})^{-1}A^{i}\partial_i{\bf{u}}^{\epsilon}
+(A^{0})^{-1}B {\bf{u}}^{\epsilon} &=&(A^{0})^{-1}{\bf F}\\ \nonumber
 {\bf{u}}^{\epsilon}(0,x)&=& \rho^{\epsilon}*\left({\bf{u}}_{0}(x)\right)
\end{eqnarray}
where $\{(\rho^{\epsilon})\}\in(0,1]$ is a family of
mollifiers. Here $\Delta_{h}$ is the Laplace-Beltrami operator on
$\Sigma$ associated with the smooth background Riemannian metric
$h_{ij}$. By adding in the second order Laplace-Beltrami terms we
obtain a system with smooth principal symbol. We may then use
classical methods of parabolic regularity theory to obtain a solution
with better analytic properties than the original hyperbolic system.

\item 
Second, we obtain the following uniform energy estimate 
\begin{equation}\label{eest}
  ||{\bf{u}}^{\epsilon}||^{2}_{L^{2}(0,T;L^{2}(\Sigma,\mathbb{R}^{N}))}
\le C\left( ||{\bf{u}}_{0}||_{L^{2}(\Sigma,\mathbb{R}^{N})}
+\int^{T}_{0}||{\bf F}(t,\cdot )||_{L^{2}(\Sigma,\mathbb{R}^{N})} dt \right).
\end{equation}
  where $C$ is independent of $\epsilon$. 
   
\item Third, we take the limit $\epsilon\rightarrow 0$ and show
  convergence in an appropriate weak sense to a regular weak solution
  as defined above.

\item Fourth, using the energy inequality (\ref{eest}) we show
  uniqueness and stability. This concludes the proof of Theorem
  \ref{tfo1}.
  
\item Fifth, we rewrite the wave equation as a symmetric hyperbolic
  problem and show that for a Lipschitz metric the corresponding $L$
  satisfies the conditions of Theorem \ref{tfo1}.
  
\item Sixth, we show that the solution of the wave equation obtained
  via the symmetric hyperbolic problem is in $H^{1}(\Sigma_{[0,T]})$. This
  concludes the proof of Theorem \ref{tfo}.
\end{enumerate}

\subsection{Approximate solutions and energy estimate}

The results we obtain make extensive use of the vanishing viscosity
method. As explained above, the first step is to show that there exist
suitable solutions to (\ref{PDE2}). This step follows directly from the
work of Evans (\cite{evans}, Th. 1 \S7.3).

\begin{Proposition}\label{espd2} {\bf( Existence of Approximate solutions)}
For each $\epsilon>0$, there exists a unique solution
  ${\bf{u}}^{\epsilon}$ of (\ref{PDE2}) with ${\bf{u}}^{\epsilon}\in
  L^{2}(0,T;H^{2}(\Sigma,\mathbb{R}^{N}))$ and ${\bf{\dot
      u}}^{\epsilon} \in L^{2}(0,T;L^{2}(\Sigma,\mathbb{R}^{N}))$.
\end{Proposition}

{\it Proof.}  
This is a variant of a standard result for parabolic
systems. Following Evans set $X=L^{\infty}(0,T;H^{1}(\Sigma,\mathbb{R}^{N}))$ and then for each ${\bf w} \in X$,
consider the linear system
\begin{eqnarray}\label{vp}
    \partial_t{\bf u}^{\epsilon}-\epsilon\Delta_h {\bf u}^{\epsilon}&=&
-(A^{0})^{-1}A^{i}\partial_i{\bf w}-(A^{0})^{-1}B{\bf w} +(A^{0})^{-1}{\bf F}\\
  {\bf u}^{\epsilon}(0,x)&=& {\bf u}_0^{\epsilon}(x) 
\end{eqnarray}
where ${\bf u}_0^{\epsilon}(x)=\rho^\epsilon(x)*({\bf u}_0(x))$.
Notice that the system is formed of $N$ scalar parabolic equations of
the form $ \partial_t\upsilon-\epsilon\Delta_{h}\upsilon=f$. The
coefficients are now all smooth and the only loss of regularity comes
from the source term on the RHS of (\ref{vp}). However as this is bounded in
$L^{2}(0,T;L^{2}(\Sigma,\mathbb{R}^{N}))$, we can apply standard results (see e.g.
\cite{evans} Th. 5 \S7.1) to show that there is a unique solution
${\bf u}^{\epsilon}\in L^{2}(0,T;H^{2}(\Sigma,\mathbb{R}^{N}))$ with
${\bf \dot u}^{\epsilon} \in L^{2}(0,T;L^{2}(\Sigma,\mathbb{R}^{N}))$.

In the same manner we can choose $\tilde{\bf w}\in X$ and find
$\tilde{\bf u}^{\epsilon}$ that solves

\begin{eqnarray}\label{vp2}
    \partial_t \tilde{\bf u}^{\epsilon}-\epsilon\Delta_h \tilde{\bf u}^{\epsilon}
&=&-(A^{0})^{-1}A^{i}\partial_i\tilde{\bf w}
-(A^{0})^{-1}B \tilde{\bf w}+(A^{0})^{-1}{\bf F}\\
\tilde{\bf u}^{\epsilon}(0,x)&=& {\bf u}_0^{\epsilon}(x)
\end{eqnarray}

Subtracting ${\bf
  u}^{\epsilon}-\tilde{\bf u}^{\epsilon}$, we find $\bar{\bf u}^{\epsilon}={\bf
  u}^{\epsilon}-\tilde{\bf u}^{\epsilon}$ solves

\begin{eqnarray}
    \partial_t\bar{\bf u}^{\epsilon}-\epsilon\Delta_h \bar{\bf u}^{\epsilon}
&=&-(A^{0})^{-1}A^{i}\partial_i\bar{\bf w}
-(A^{0})^{-1}B \bar{\bf w}\\
   \bar{\bf u}^{\epsilon}(0,x)&=&0
\end{eqnarray}
where $\bar{\bf w}={\bf w}-\tilde{\bf w}$.
Using standard energy estimates for solutions of parabolic
equations we have that $\bar{\bf u}$ satisfies:

\begin{eqnarray}\nonumber
 &\displaystyle\sup_{0\le t\le T}||\bar{\bf{ u}}^{\epsilon}(t)||^{2}_{H^{1}(\Sigma,\mathbb{R}^{N})}&\\ \nonumber
    &\le C(T,\epsilon)\left(||(A^{0})^{-1}(A^{i}\partial_i\bar{\bf w}+B \bar{\bf w})||^{2}_{L^{2}(0,T;L^{2}(\Sigma,\mathbb{R}^{N}))}\right)&\\
  &\le C(T,\epsilon)\left({\displaystyle \sup_{0\le t\le T}}||\bar{\bf w}(t)||^{2}_{H^{1}(\Sigma,\mathbb{R}^{N})}\right) &
\end{eqnarray}
\noindent
Thus
\begin{equation}
  ||\bar{\bf u}^{\epsilon}||_{L^{\infty}(0,T; (\Sigma,\mathbb{R}^{N}))}\le  C(T,\epsilon)||\bar{\bf w}||_{L^{\infty}(0,T; (\Sigma,\mathbb{R}^{N}))} 
\end{equation}
\noindent
Therefore, if $T$ is small enough such that
$C(T,\epsilon)\le\frac{1}{2}$ we obtain that

\begin{equation}\label{hb}
  ||\bar{\bf u}^{\epsilon}||_{L^{\infty}(0,T; (\Sigma,\mathbb{R}^{N}))}\le\frac{1}{2} ||\bar{\bf w}||_{L^{\infty}(0,T; (\Sigma,\mathbb{R}^{N}))} 
\end{equation}
so that
\begin{equation}\label{hb2}
  ||{\bf u}^{\epsilon}-\tilde{\bf u}^{\epsilon}||_{L^{\infty}(0,T; (\Sigma,\mathbb{R}^{N}))}\le\frac{1}{2} ||{\bf w}-\tilde{\bf w}||_{L^{\infty}(0,T; (\Sigma,\mathbb{R}^{N}))} 
\end{equation}
This implies that we have a contraction mapping and the hypothesis of
Banach's fixed point theorem is satisfied for the mapping
\begin{eqnarray*}
M:L^{\infty}(0,T;(\Sigma,\mathbb{R}^{N}))&\rightarrow&
L^{\infty}(0,T;(\Sigma,\mathbb{R}^{N}))\\
{\bf w} & \mapsto & {\bf u}^{\epsilon}
\end{eqnarray*}
which therefore has a unique fixed point which solves (\ref{PDE2}).
\noindent
If $C(T,\epsilon) > \frac{1}{2} $ we can choose $T_{1}$ small enough
such that $C(T_{1},\epsilon)\le\frac{1}{2}$ and then repeat the above argument
for intervals $[0,T_{1}],[T_{1},2T_{1}],...,[nT_{1},T]$.
\noindent
In either case we obtain a solution ${\bf{u}}^{\epsilon}$ which solves
(\ref{PDE2}) on the interval $[0,T]$. Standard parabolic regularity
theory (see e.g.  Th. 5 \S7.1 \cite{evans}) then gives us 
${\bf{u}}^{\epsilon}\in L^{2}(0,T;H^{2}(\Sigma,\mathbb{R}^{N}))$ and that
${\bf \dot u}^{\epsilon} \in L^{2}(0,T;L^{2}(\Sigma,\mathbb{R}^{N}))$, 
which concludes the proof of Theorem \ref{espd2}.
$\boxdot$\\
Note that Evans goes on to use Th. 6 \S7.1 \cite{evans} to obtain an
improved regularity result showing that ${\bf{u}}^{\epsilon}\in
L^{2}(0,T;H^{3}(\Sigma,\mathbb{R}^{N}))$ and that the time derivative ${\bf{\dot u}}^{\epsilon}
\in L^{2}(0,T;H^{1}(\Sigma,\mathbb{R}^{N}))$. However we do not need
this result.

\subsection{Energy estimates}

The next step is to obtain the uniform energy estimate in $\epsilon$
for the solutions ${\bf{u}}^{\epsilon}$. This is the content of the
following proposition.
\vspace{5mm}
\begin{Proposition}\label{ee}
There exists a constant $C$ depending on $T$, $\Sigma$, $h_{ij}$, $\partial_{l}h_{ij},$ \\
 $\displaystyle \sup_{x\in \Sigma_{[0,T]}}\left(|(A^{0})^{-1}|,|A^{i}|,|B|,|\partial_{t}(A^{0})^{-1}|,|\partial_{t}A^{i}|,|\partial_{j}(A^{0})^{-1}|,|\partial_{j}A^{i}|\right)$  such that

\begin{equation}\label{estimate}
  ||{\bf{u}}^{\epsilon}||^{2}_{L^{2}(0,T;L^{2}(\Sigma,\mathbb{R}^{N}))}
  \le C\left( ||{\bf{u}}_{0}||_{L^{2}(\Sigma,\mathbb{R}^{N})}+\int^{T}_{0}||{\bf{F}}(t,\cdot )||_{L^{2}(\Sigma,\mathbb{R}^{N})} dt \right)
\end{equation}
\noindent
Therefore the estimate is independent of $\epsilon$.

\end{Proposition}
\noindent
{\it{Proof of Proposition \ref{ee}.}}
Taking the time derivative of $||{\bf u}^{\epsilon}||^2_{\ellv}$ gives:
\begin{equation}\label{sh1}
\begin{split}
  &\frac{d}{dt}\left(||{\bf{u}}^{\epsilon}(t)||^{2}_{\ellv}\right)=2\left({\bf{u}}^{\epsilon}, {\bf \dot u}^{\epsilon}\right)_{\ellv}\\
  &=2\left({\bf{u}}^{\epsilon},\epsilon\Delta_h {\bf{u}}^{\epsilon}-(A^{0})^{-1}A^{i}\partial_i{\bf{u}}^{\epsilon}-(A^{0})^{-1}B{\bf{u}}^{\epsilon}+(A^{0})^{-1}{\bf F}\right)_{\ellv}
  \end{split}
\end{equation}

\noindent

We have estimates for the following terms in (\ref{sh1})
\begin{eqnarray}\label{sh2}
  |({\bf{u}}^{\epsilon},(A^{0})^{-1}{\bf F})_{\ellv}|&\le&C_{1} (||{\bf{u}}^{\epsilon}(t)||^{2}_{\ellv}+||{\bf{F}}||^{2}_{\ellv})\\ \label{sh3}
  ({\bf{u}}^{\epsilon},\epsilon\Delta_h {\bf{u}}^{\epsilon})_{\ellv}&=&-\epsilon\displaystyle\sum^{N}_{l=1}\int_{\Sigma}h^{ij}\partial_{i}(u^{l})\partial_{j}(u^{l})\nu_{h}\le 0\\\label{sh4}
  \left|({\bf{u}}^{\epsilon},(A^{0})^{-1}B {\bf{u}}^{\epsilon})\right|&\le& C_{2} || {\bf{u}}^{\epsilon}||^{2}_{\ellv}.
\end{eqnarray}
However we also require an estimate for $({\bf u}^{\epsilon},
(A^0)^{-1}A^i\partial_i{\bf u}^{\epsilon})_{\ellv}$, which is the
remaining term in (\ref{sh1}).  This term can be estimated by
applying a suitable integration by parts.

We first assume that the smooth background Riemannian metric is
$h_{ij}=\delta_{ij}$, write $(A^0)^{-1}A^i$ as $\tilde A^i$ and
  estimate $({\bf u}^{\epsilon}, \tilde A^i\partial_i{\bf
    u}^{\epsilon})_{\ellv}$. Using the fact
  that the $\tilde A^i$ are symmetric, we then have

\begin{eqnarray}
&({\bf u}^{\epsilon}, \tilde A^i\partial_i{\bf u}^{\epsilon})_{\ellv}\\
=& \int_\Sigma{\bf u}^{\epsilon}\bcdot\left(\tilde A^i\partial_i{\bf u}^{\epsilon}\right) d^nx \\
=&\frac{1}{2}\int_\Sigma\partial_i\left(\tilde A^i{\bf u}^{\epsilon}\right)\bcdot {\bf u}^{\epsilon}d^3x -\frac{1}{2}\int_\Sigma\left(\partial_i\tilde A^i\right){\bf u}^{\epsilon}\bcdot {\bf u}^{\epsilon}d^nx \\
=& -\frac{1}{2}\int_\Sigma\left(\partial_i\tilde A^i\right){\bf u}^{\epsilon}\bcdot {\bf u}^{\epsilon}d^3x 
\end{eqnarray}
So that 
\begin{eqnarray}
|({\bf u}^{\epsilon}, \tilde A^i\partial_i{\bf u}^{\epsilon})_{\ellv}| &\leqslant& 
\frac{1}{2} \left|\int_\Sigma\left(\partial_i\tilde A^i\right){\bf u}^{\epsilon}\bcdot {\bf u}^{\epsilon}d^nx\right|\\
& \leqslant & C_{3}||{\bf u}^{\epsilon}||^2_{\ellv}
\end{eqnarray}
where the constant $C_{3}$ is independent of $\epsilon$ and we have used
the fact that $\partial_i \tilde A^i$ is bounded. In the case of a general
background Riemannian metric $h_{ij}$ rather than the ordinary
divergence of $\tilde A^i$, one obtains the divergence with respect to
the background metric $h_{ij}$ and the corresponding result is
\begin{eqnarray}\label{sh5}
|({\bf v}, \tilde A^i\partial_i{\bf v})_{\ellv}| &\leqslant& 
\frac{1}{2} \left|\int_\Sigma\left((\partial_i+\tilde{\Gamma}^k_{ki})\tilde A^i\right){\bf u}^{\epsilon}\bcdot {\bf u}^{\epsilon}\nu_h\right|\\
& \leqslant & C_4||{\bf u}^{\epsilon}||^2_{\ellv}
\end{eqnarray}
where again $C_4$ is independent of $\epsilon$. For the case where
$\Sigma$ is an open, bounded set of $\mathbb{R}^{n}$ with smooth
boundary $\partial \Sigma$, one needs a slightly more complicated
argument where one approximates ${\bf u}^{\epsilon}$ by smooth functions of compact
support (see \cite{evans} \S7.3 for details).

Using all the available estimates (\ref{sh2}), (\ref{sh3}), (\ref{sh4})
and (\ref{sh5}) in (\ref{sh1}) we obtain the estimate

\begin{equation}
 \frac{d}{dt}\left(||{\bf{u}}^{\epsilon}(t)||^{2}_{\ellv}\right)\le C_{5} \left(|| {\bf{u}}^{\epsilon}||^{2}_{\ellv}+|| {\bf{F}}||^{2}_{\ellv}\right)
\end{equation}
\noindent
Using Gronwalls's inequality and the fact that

\begin{equation}
   ||{\bf{u}}^{\epsilon}(0,x)||_{\ellv}\le||{\bf{u}}_{0}(x)||_{\ellv}
\end{equation}
we obtain the estimate

\begin{equation}\label{eepd2}
\displaystyle \sup_{0\le t\le T} ||{\bf{u}}^{\epsilon}(t)||^{2}_{\ellv}\le C_{6} \left(||{\bf{u}}_{0} ||^{2}_{\ellv}  +\int_{0}^{T}|| {\bf{F}}||^{2}_{\ellv}dt\right)
\end{equation}
Finally noting that 
\begin{equation}\label{l2li}
  ||{\bf{u}}^{\epsilon}||^{2}_{L^{2}(0,T;L^{2}(\Sigma,\mathbb{R}^{N}))}\le T {\displaystyle \sup_{0\le t\le T}}||{\bf{u}}^{\epsilon}(t)||^{2}_{\ellv}
\end{equation}
and using this in the estimate (\ref{eepd2}) we obtain (\ref{estimate}) which concludes the proof.
$\boxdot$ \\

\subsection{Existence, Uniqueness and Stability}

In this section we show the existence and uniqueness of the initial
value problem.  In Proposition \ref{espd2} we obtained solutions ${\bf u}^{\epsilon}$ in
$L^{2}(0,T;H^{2}(\Sigma,\mathbb{R}^{N}))$ to the parabolic system
(\ref{PDE2}). Using the Banach–Alaoglu Theorem there exists a
subsequence $\{{\bf{u}}^{\epsilon_{k}}\}^{\infty}_{k=1}$ that
converges weakly to a function ${\bf{u}}$ in $L^{2}(0,T;L^{2}(\Sigma,\mathbb{R^{N}}))$.
 We now show that this converges to a weak solution of (\ref{PDE1}).
\noindent

First we choose a function ${\bf{\tilde{w}}} \in
\tilde{H}^{2}({\Sigma_{[0,T]}},\mathbb{R}^{N})$, take the dot
product with equation (\ref{PDE2}) and integrate over $t$ and
$x$. This gives

\begin{equation}\label{old}
\begin{split}
   &\int_{0}^{T}((A^0)^{-1}L{\bf{u}}^{\epsilon}, {\bf{\tilde{w}}})_{\ellv}-\epsilon (\Delta_h{\bf{u}}^{\epsilon},{\bf{\tilde{w}}})_{\ellv}dt\\=& \int_{0}^{T}((A^{0})^{-1} {\bf F},  {\bf{\tilde{w}}})_{\ellv}dt
\\
   \end{split}
\end{equation}

\noindent

Integrating by parts we obtain 
\begin{equation}\label{id}
\begin{split}
   &\int_{0}^{T}({\bf{u}}^{\epsilon}, \tilde{L} {\bf{\tilde{w}}})_{\ellv}-\epsilon ({\bf{u}}^{\epsilon},\Delta_{h} {\bf{\tilde{w}}})_{\ellv}dt\\=& \int_{0}^{T}((A^{0})^{-1}{\bf F},  {\bf{\tilde{w}}})_{\ellv}dt+({\bf{u}}^{\epsilon}(0),  {\bf{\tilde{w}}}(0))_{\ellv}\\
   \end{split}
\end{equation}
where the operator $\tilde L$ is defined by 
\begin{equation}
  \tilde{L} {\bf{\tilde{w}}}\coloneqq - \partial_t{\bf{\tilde{w}}}-\partial_i(A^{i}(A^{0})^{-1}{\bf{\tilde{w}}})+B^{T}((A^{0})^{-1}) {\bf{\tilde{w}}} -\tilde{\Gamma}^{l}_{il}A^{i}(A^{0})^{-1}  {\bf{\tilde{w}}}
\end{equation}
\noindent
Then taking the limit $k \to \infty$ and using the weak convergence of ${\bf{u}}^{\epsilon_k} \rightharpoonup {\bf{u}}$ and that ${\bf{u}}^{\epsilon_k}(0)\rightarrow (A^{0})^{-1}(0){\bf{u}}_{0}$ in $L^{2}(\Sigma,\mathbb{R}^{N})$ we obtain

\begin{equation}\label{wsol2}
\begin{split}
   &\int_{0}^{T}({\bf{u}}, \tilde{L} {\bf{\tilde{w}}})_{\ellv}dt\\=& \int_{0}^{T}((A^{0})^{-1}{\bf F},  {\bf{\tilde{w}}})_{\ellv}dt+({\bf{u}}_{0},  {\bf{\tilde{w}}}(0))_{\ellv}\\
   \end{split}
\end{equation}

The above equality was obtained for $ {\bf{\tilde{w}}} \in
\tilde{H}^{2}({\Sigma_{[0,T]}},\mathbb{R}^{N})$, however the equation remains
well-defined for $ {\bf{\tilde{w}}} \in
\tilde{H}^{1}({\Sigma_{[0,T]}},\mathbb{R}^{N})\subset
C^{0}([0,T],L^{2}(\Sigma))$ . We now show that not only is the
equation well-defined, but it remains valid for $ {\bf{\tilde{w}}} \in
\tilde{H}^{1}({\Sigma_{[0,T]}},\mathbb{R}^{N})$. The method involves taking the convolution with a family of mollifiers $\rho^\delta$ and then passing to the limit $\delta\rightarrow 0$.

Let $ {\bf{\tilde{w}}} \in \tilde{H}^{1}({\Sigma_{[0,T]}},\mathbb{R}^{N})$ and define ${\bf{\tilde{w}}}^{\delta}=\rho^{\delta}* {\bf{\tilde{w}}}$
where we have chosen $\delta$ close enough to zero such that
${\bf{\tilde{w}}}^{\delta}(T,\cdot)=0$. Therefore we have

\begin{equation}
\begin{split}
   &\int_{0}^{T}({\bf{u}}, \tilde{L} {\bf{\tilde{w}}}^{\delta})_{\ellv}dt\\=& \int_{0}^{T}((A^{0})^{-1} {\bf F},  {\bf{\tilde{w}}}^{\delta})_{\ellv}dt+({\bf{u}}(0),  {\bf{\tilde{w}}}^{\delta}(0))_{\ellv}\\
   \end{split}
\end{equation}

Taking the limit $\delta\rightarrow 0$ and using the Schwartz inequality
we have the following limits

\begin{eqnarray}
   {\bf{\tilde{w}}}^{\delta}_{t}&\rightarrow& {\bf{\tilde{w}}}_{t} \text{ in } L^{2}(0,T; L^{2}(\Sigma,\mathbb{R}^{N}))\\
   -(A^{i}(A^{0})^{-1}){\bf{\tilde{w}}}^{\delta}_{i}&\rightarrow& -(A^{i}(A^{0})^{-1}){\bf{\tilde{w}}}_{i}\text{ in } L^{2}(0,T; L^{2}(\Sigma,\mathbb{R}^{N}))\\
  B^{T}((A^{0})^{-1})  {\bf{\tilde{w}}}^{\delta}&\rightarrow& B^{T}((A^{0})^{-1}) {\bf{\tilde{w}}} \text{ in } L^{2}(0,T; L^{2}(\Sigma,\mathbb{R}^{N}))\\
   \tilde{\Gamma}^{k}_{lk}A^{l}(A^{0})^{-1} {\bf{\tilde{w}}}^{\delta}&\rightarrow& \tilde{\Gamma}^{k}_{lk}A^{l}(A^{0})^{-1} {\bf{\tilde{w}}} \text{ in } L^{2}(0,T; L^{2}(\Sigma,\mathbb{R}^{N}))\\
   {\bf{\tilde{w}}}^{\delta}(0)&\rightarrow& {\bf{\tilde{w}}}(0) \text{ in } L^{2}(\Sigma,\mathbb{R}^{N})
\end{eqnarray}

We therefore conclude that

\begin{equation}\label{wsol3}
\begin{split}
   &\int_{0}^{T}({\bf{u}}, \tilde{L} {\bf{\tilde{w}}})_{\ellv}dt\\=& \int_{0}^{T}((A^{0})^{-1} {\bf F},  {\bf{\tilde{w}}})_{\ellv}dt+({\bf{u}}_{0},  {\bf{\tilde{w}}}(0))_{\ellv}\\
   \end{split}
\end{equation}
for $ {\bf{\tilde{w}}} \in \tilde{H}^{1}({\Sigma_{[0,T]}},\mathbb{R}^{N})$    \\


If we now take $ {\bf{{w}}}\in
C^{\infty}({\Sigma_{[0,T]}},{\mathbb{R}^{N}})$, with $\supp(
{\bf{{w}}})\subseteq [0,T)\times\Sigma$, and multiply it by $A^{0}$
we obtain that $A^{0} {\bf{{w}}}\in
\tilde{W}^{1,\infty}(\Sigma_{[0,T]},\mathbb{R}^{N}) \subset \tilde{H}^1(\Sigma_{[0,T]},\Real^N)$ and  $\supp(A^0\bf{w})\subset \supp(\bf{w})$ .

 We may therefore insert ${\bf \tilde w}=A^0{\bf w}$ in
(\ref{wsol3}) which gives

 \begin{equation}
 \begin{split}
   &\int_{0}^{T}({\bf{u}}, \tilde{L}A^{0} {\bf{{w}}})_{\ellv}dt\\=& \int_{0}^{T}((A^{0})^{-1} {\bf F}, (A^{0}) {\bf{{w}}})_{\ellv}dt+\left({\bf{u}}_{0}, (A^{0}){\bf{{w}}}|_{t=0}\right)_{\ellv}
   \end{split}
\end{equation}
\noindent
which can be rewritten as

 \begin{equation}
   \int_{0}^{T}({\bf{u}}, L^{*} {\bf{{w}}})_{\ellv}dt= \int_{0}^{T}({\bf F},  {\bf{{w}}})_{\ellv}dt+\left((A^{0}(0)){\bf{u}}_{0},  {\bf{{w}}}(0)\right)_{\ellv}
\end{equation}
for all $ {\bf{{w}}}\in C^{\infty}({\Sigma_{[0,T]}},{\mathbb{R}^{N}})$
with $\supp( {\bf{{w}}})\subseteq [0,T)\times\Sigma$, where $L^{*}$
is the formal adjoint defined by equation
(\ref{adjoint}).  

We have therefore proved that the ${\bf{u}}$ obtained by taking the
limit of the subsequence $\{{\bf{u}}^{\epsilon_{k}}\}^{\infty}_{k=1}$
is a weak solution lying in $L^{2}(0,T;L^{2}(\Sigma,\mathbb{R^{N}}))$ 
with initial data ${\bf{u}}_{0}$ as in Definition \ref{ddeeff}. \\

As the norm function is lower semi continuous, we may take the limit of
equation (\ref{estimate}) to obtain the estimate
\begin{eqnarray}\label{eup}
  ||{\bf{u}}||^{2}_{L^{2}(0,T;L^{2}(\Sigma,\mathbb{R}^{N}))}&\le& \lim_{{k}\rightarrow \infty}||{\bf{u}}^{\epsilon_{k}}||^{2}_{L^{2}(0,T;L^{2}(\Sigma,\mathbb{R}^{N}))}\\
  &\le& C \left(||{\bf{u}}_{0} ||^{2}_{\ellv}  +\int_{0}^{T}||{\bf F}||^{2}_{\ellv}dt\right)
\end{eqnarray}
So that the solution we have obtained is a regular weak solution. \\

To show uniqueness we consider two functions ${\bf{u}}_{1},{\bf{u}}_{2}$
which are both regular weak solutions. Then
${\bf{u}}={\bf{u}}_{1}-{\bf{u}}_{2}$ is a regular weak solution with source
${\bf F}=0$ and initial data ${\bf{u}}_{0}=0$. Moreover the solution
satisfies the energy estimate (\ref{eeE}) as shown above. Therefore
 \begin{equation}
  ||{\bf{u}}||_{L^{2}(0,T; L^{2}(\Sigma,\mathbb{R}^{N}))} =0
  \end{equation}
This implies ${\bf{u}}=0$ and therefore ${\bf{u}}_{1}={\bf{u}}_{2}$.\\

The final step in establishing well-posedness is to prove the
stability of the solution with respect to initial data. To make the
concept precise we say that the solution ${\bf{u}}$ is continuously
stable in $L^{2}(0,T;L^{2}(\Sigma,\mathbb{R}^{N}))$ with respect to initial data
${\bf{u}}_{0}$ in $L^{2}(\Sigma,\mathbb{R}^{N})$ , if given $\epsilon >0$ there is a $\delta$ depending on
${\bf{u}}_{0}$ such that if ${\bf \tilde u}_0 \in
L^{2}(\Sigma,\mathbb{R}^{N})) $ with:
\begin{equation}
  \Arrowvert {\bf \tilde {u}}_{0}-{{\bf{u}}_{0}}\Arrowvert_{\ellv}\le \delta,
\end{equation}
then the corresponding weak solution ${\bf \tilde u}$ with source function ${\bf F}$ satisfies 
\begin{equation}
  \Arrowvert {\bf{\tilde{u}}}-{\bf{{u}}}\Arrowvert_{L^{2}(0,T;L^{2}(\Sigma,\mathbb{R}^{N}))}\le \epsilon
\end{equation} 

The stability results follows from using the energy inequality shown in Theorem \ref{ee} for the difference  ${\bf{\tilde{u}}}-{\bf{{u}}}$ which gives
\begin{eqnarray}
  \Arrowvert {\bf{\tilde{u}}}-{\bf{{u}}}\Arrowvert^{2}_{L^{2}(0,T;L^{2}(\Sigma,\mathbb{R}^{N}))}&\le& C  ||\tilde{{\bf{u}}_{0}}-{\bf{u}}_{0}||^{2}_{\ellv}
\end{eqnarray}

Now choosing $\delta=\frac{\epsilon}{C}$ we obtain the inequality:

\begin{equation}
   \Arrowvert {\bf{\tilde{u}}}-{\bf{{u}}}\Arrowvert_{L^{2}(0,T;L^{2}(\Sigma,\mathbb{R}^{N}))}^{2} \leq \epsilon^{2}
\end{equation}
which establishes stability with respect to the initial data.

This concludes the proof of Theorem \ref{tfo1}.

\subsection{The wave equation}

In order to apply the results of Theorem 1 to applications in general
relativity we will show here how the wave equation can be written as a first
order linear symmetric problem.

We define
\begin{equation}\label{V}
  {\bf{{v}}}=(\partial_{1}u,...,\partial_{n}u,\partial_{t}u,u)^{T}
   =(v^1,...,v^n,v^{n+1},v^{n+2})^{T} \in \Real^{n+2}
\end{equation}
\noindent
and the  symmetric $(n+2)\times(n+2)$ matrices $A^\mu$ 
by: \\
\vskip1mm
$
A^{0} =
 \begin{pmatrix}
  g^{11} & g^{12} & \cdots & 0& 0 \\
  g^{21} & g^{22} & \cdots & 0&0 \\
  \vdots  & \vdots  & \ddots & \vdots &\vdots \\
 0 & 0 & \cdots &-g^{00} &  0\\
  0 & 0 & \cdots&0&  1
   \end{pmatrix}$\\
\vskip2mm

$
A^{k} =
 \begin{pmatrix}
  0 & 0 & \cdots &g^{1k}& 0 \\
  0 & 0 & \cdots & g^{2k}&0 \\
  \vdots  & \vdots  & \ddots & \vdots &\vdots \\
 g^{1k} &g^{2k} & \cdots &2 g^{0k} &  0\\
  0 & 0 & \cdots&0&  0
 \end{pmatrix}$.\\
 \noindent
We further define the matrix $B$ to be given by\\

$B =
 \begin{pmatrix}
  0 & 0 & \cdots &0& 0 \\
  0 & 0 & \cdots & 0&0 \\
  \vdots  & \vdots  & \ddots & \vdots &0\\
 g^{ab}\Gamma^{1}_{ab} &g^{ab}\Gamma^{2}_{ab} & \cdots & g^{ab}\Gamma^{0}_{ab} &  -m^{2} \\
  0 & 0 & \cdots&-1&  0
\end{pmatrix}.\\
$ \\
where $\Gamma^c_{ab}$ are the connection coefficients of the
spacetime metric $g_{a b}$.  
\noindent
We also define the $\Real^N$-valued vector function ${\bf F}$ by  ${\bf F}=(0,0,\dots,-f,0)^T$.

\noindent
In this way, we may rewrite the scalar wave equation (\ref{lpde}) as a
first order system which has the form
\begin{eqnarray}\label{ls}
L{\bf{{v}}}=  A^{0}\partial_{t}{\bf{{v}}}-A^{i}\partial_{i}{\bf{{v}}}+B{\bf{{v}}}&=&{\bf F}\\
  {\bf{{v}}}(0,\cdot)&=&{\bf{{v}}}_{0}(\cdot)
\end{eqnarray}

We may then use the Theorem \ref{tfo1} to establish well-posedness of (\ref{ls}). To prove existence of solutions to the wave equation, we
therefore need to prove that the solution ${\bf{{v}}}$ of the symmetric
hyperbolic system (\ref{ls}) has the form $
{\bf{{v}}}=(\partial_{1}u,...,\partial_{n}u,\partial_{t}u,u)^{T}$. 

 We now mollify our solution ${\bf{{v}}}$ using a strict delta net to obtain
 a sequence of smooth functions
 ${\bf{{v}}}^{\epsilon}=\rho^{\epsilon}*{\bf{{v}}}=(v^{1}_{\epsilon},...,v^{n+2}_{\epsilon})^{T}$
 that satisfies:
 
$$\int^{T}_{0}(L{\bf{{v}}}^{\epsilon},{\bf w})_{\ellv}dt=\int^{T}_{0}({\bf F}^{\epsilon}, {\bf w})_{\ellv}dt$$
\noindent

for a suitable mollified ${\bf F}^{\epsilon}$ of the form $(0,0,\dots,-h^{\epsilon},0)^T$ where $h^{\epsilon}$ is any $L^{2}$ function.

Then 
$$\int^{T}_{0}(L{\bf{{v}}}^{\epsilon}-{\bf F}^{\epsilon},{\bf w})_{\ellv}dt=0$$
\noindent
for all ${\bf w}\in C^{\infty}({\Sigma_{[0,T]}},{\mathbb{R}^{N}})$ and therefore $L{\bf{{v}}}^{\epsilon}={\bf F}^{\epsilon}$ almost everywhere so we obtain that:

$$\partial_{t}v^{j}_{\epsilon}=\partial_{j}v^{n+1}_{\epsilon}$$ and 

$$\partial_{t}v^{n+2}_{\epsilon}=v^{n+1}_{\epsilon}.$$

\noindent
Also, we choose initial data such
that
\begin{equation}\label{idc}
  \partial_{i}v^{n+2}_{\epsilon}(0,\cdot)=v^{i}_{\epsilon}(0,\cdot), \qquad i=1,2,\dots,n
\end{equation}
 
\noindent
We now define
$u_{\epsilon}=v^{n+2}_{\epsilon}=\rho^{\epsilon}*v^{n+2}$ and obtain

\begin{eqnarray}
  \partial_{i}u_{\epsilon}(\tau)&=&v^{i}_{\epsilon}(0,\cdot)+\int^{\tau}_{0}\partial_{i}\partial_{t}v^{n+2}_{\epsilon}dt\\
  &=&v^{i}_{\epsilon}(0,\cdot)+\int^{\tau}_{0}\partial_{t}v^{i}_{\epsilon}dt\\
  &=&v^{i}_{\epsilon}(\tau)
\end{eqnarray}

Taking now into account that ${\bf{{v}}}^{\epsilon}\rightarrow {\bf{{v}}}$ in
$L^{2}([0,T],L^{2}(\Sigma,\mathbb{R}^{n+2}))$ and that the convolution and
derivatives commute we obtain the result that ${\bf{{v}}}$ is an
$L^{2}([0,T],L^{2}(\Sigma,\mathbb{R}^{n+2}))$ function of the form
$(\partial_{1}u,...,\partial_{n}u,\partial_{t}u,u)$.

Collecting the results from this section we have established Theorem \ref{tfo}.\\

\section{Applications}\label{Discussion}

  Although in the following three examples the spacetimes are not
  spatially compact we will assume we are working in a local region of
  the form $\Sigma_{[0,T]}=[0,T]\times \Sigma$ as described in the geometric setting. \\
 \noindent

{\bf{Junction Conditions}}
\noindent
There is a precise mathematical formalism proposed by Israel to
describe the junction conditions for two regular spacetimes joined
along a non-null singular hypersurface $\Lambda$ \cite{thin}. He noted
that if we consider two half-spaces $V^+$ and $V^-$, a singular
hypersurface, $\Lambda$, can be fully characterised by the different
extrinsic curvatures (second fundamental forms) associated with its
embeddings in $V^+$ and $V^-$ and a continuous matching condition of the metric through the common boundary. If we use Gaussian coordinates based on
$\Lambda$, then the normal derivatives of the metric have a jump
across $\Lambda$ with the metric being continuous along $\Lambda$.
This scenario satisfies the analytic conditions required for the application of Theorem
\ref{tfo} to apply. Notice however that the theorem does not need to
make assumptions on the time dependence or matter content of the spacetime. \\

\noindent
{\bf{Impulsive Gravitational Waves}}
\noindent
A spacetime that contains impulsive gravitational waves described in
double null coordinates has line element given by:

\begin{equation}
  ds^{2}=2dudv-(1-u\Theta(u))^{2}dy^{2}-(1+u\Theta(u))^{2}dz^{2}
\end{equation} 
\noindent
where $\Theta(u)$ is the Heaviside step function.  The spacetime is
vacuum, but has a Weyl tensor with delta function components
\begin{eqnarray}
C_{uyuy}&=&-\delta(u)\nonumber \\\nonumber
C_{uzuz}&=& \delta(u) 
\end{eqnarray}

It is important to notice that although the curvature is not bounded this condition is
not relevant for Theorem \ref{tfo} to apply. \\

\noindent
{\bf{Brane-world Cosmologies}} 
\noindent
The Randall–Sundrum Brane-Worlds (RS) are models that explore gravity
beyond classical general relativity \cite{maeda}, and
also appear in a cosmological context \cite{strin}.  In the RS model
in $ AdS_{5}$ one has that in Gaussian normal coordinates
$X^{A}=(t, x^{i},y)$ based on the brane at $y=0$, the model has the line
element
\begin{equation}
   ds^{2}=e^{-2|y|/L}(-dt^{2}+ dx^{i2})+dy^{2}
\end{equation}

 This spacetime again satisfies the conditions for the Theorem
\ref{tfo} to be applicable and therefore solutions with finite
energy
 exist.
 
 Notice that the well-posedness result can be
extended to other brane
 world models and even collision of branes as
long as the spacetime
 satisfies the assumptions of Theorem
\ref{tfo}. Therefore, one can
 consider that dynamical models of
colliding branes (see e.g. \cite{bla1}) do not produce strong
gravitational singularities provided that the spacetime remains
$C^{0,1}$ during all the process.

\subsection{Main result and the relation to previous work}

  The main result of the paper (Theorem \ref{tfo}) shows that spacetimes which are usually
  thought of as singular may be regarded as regular if one adopts the
  point of view that true singularities make the local dynamics of
  test fields ill-defined. We have established general conditions
  under which linear wave equations are locally well-posed in
  spacetimes with weak singularities where the singularity is
  concentrated on a submanifold. In particular, the results can be
  applied to spacetimes with shell-crossing singularities, surface
  layers and hypersurface singularities  of regularity $C^{0,1}$. 

   We establish local well-posedness for general first
  order linear symmetric hyperbolic systems with coefficients with low
  regularity.  We show that unique stable solutions exist in
  $L^{2}(0,T; L^{2}(\Sigma, \mathbb{R}^{N}))$. This solution
  corresponds in the second order formalism to a finite energy
  solution in $H^{1}$ of the wave equation. Moreover, the main advantage in writing
  the problem as a first order system is that the existence of a
  covariantly constant timelike vector fields and the condition on the
  curvature are not needed which are key conditions in previous works \cite{conical, VW}.
   Therefore, the results obtained extend previous results of Vickers and Wilson \cite{VW}
   and Ishibashi and Hosoya \cite{IH} by allowing a larger
  class of non-vacuum time dependant spacetimes. We also establish not
  only the existence and uniqueness of solutions but also their
  stability and local well posedness.

\end{section}  

\section*{Acknowledgements}

The authors would like to thank CONACyT for supporting this work through a CONACyT Graduate Fellowship. 


\addcontentsline{toc}{chapter}{References}
\begin{thebibliography}{99}

\bibitem{leray} J. Leray {\it{``Hyperbolic Differential Equations''}}, Institute for Advanced Study,  (1953) 

\bibitem{bernal} A. Bernal, M. Sanchez   {\it{Class. Quantum Grav.}} {\bf{24}} 745 2007

\bibitem{hawking} S.W. Hawking, G.F.R. Ellis , {\it{The large scale structure of spacetime}}  Cambridge University Press, (1974).

\bibitem{singc11} M. Kunzinger, R. Steinbauer, M. Stojkovic, J. Vickers  {\it{Class. Quantum Grav.} }{\bf{32}} 075012  2015
\bibitem{cm}P.T. Chru\'sciel, J.D.E. Grant  {\it{Class. Quantum Grav.}} {\bf{29}} 145001 2012

\bibitem{generalized} C. J. S. Clarke {\it{Class. Quantum Grav.}} {\bf{15}} 975  1998 
\bibitem{thin} W. Israel {\it{Il Nuovo Cimento}} B Series 10 11 Luglio , Volume 44, Issue 1, 1-14 1966


\bibitem{penrose} R. Penrose The geometry of impulsive gravitational waves. In L. O’Raifeartaigh, editor, {\it{General Relativity}}, pages 101–115. Clarendon Press, (1972).



\bibitem{Clarke&odonnell} C. J. S. Clark and N O'Donnell {\it{Rendiconti del Seminario Matematico, Universit\'a Torino}} {\bf 50} 39 1992

\bibitem{part}I. R\'acz  {\it{Class. Quantum Grav.}} {\bf{27}} 155007 2010


\bibitem{wald} R. Wald  {\it{ J Math Phys.}} {\bf 21}, 2820 1980

\bibitem{Marolf} G. Horowitz, D. Marolf  {\it{Phy. Rev. D}} {\bf 52},  5670 1995

\bibitem{kay} B.S. Kay, U.M. Studer  {\it{Communications in Mathematical Physics}} {\bf 139}, 103 1991


\bibitem{IH} A. Ishibashi, A. Hosoya  {\it{Phy. Rev. D}} {\bf 60},  104028 1999

\bibitem{ys2} Y. Sanchez Sanchez, J. A. Vickers 
\emph{Class.Quant.Grav.} {\bf{33}} 205002 2016 
\bibitem{GMS09} J.D.E. Grant, E. Mayerhofer, and R. Steinbauer, \emph{Comm. Math. Phys.} {\bf 285} 399 2009 

\bibitem{evans} L.C. Evans {\it{Partial Differential Equations}}
  American Mathematical Society, (2002).


\bibitem{VW} J. A. Vickers, J. P. Wilson  {\it{ArXiv:0101018}} 2001


\bibitem{ringstrom} H. Ringstr\"om {\it{ The Cauchy Problem in General Relativity}} ESI Lectures in Mathematics and Physics (2009)


  
\bibitem{cross} P. Szekeres, A. Lun  {\it{J. Austral. Math. Soc. Ser. B}}  {\bf{41}}, 167-179  1999

\bibitem{gt} R. Geroch, J. Traschen {\it{Phys. Rev. D}} {\bf{36}}, 1017  1987
\bibitem{gerocht} R. Steinbauer,  J.A.G Vickers  {\it{Class. Quantum Grav.}}   {\bf{26}} 6 2008


{\bibitem{pams} A. Almheiri, D. Marolf, J. Polchinski, J. Sully  {\it{Journal of High Energy Physics}} {\bf{2}} 62} 2013

\bibitem{SS} Y. Sanchez Sanchez  { \it{General Relativity and Gravitation}} {\bf{47}} 80 2015

  

\bibitem{fried} K.O. Friedrichs  { \it{Comm. Pure Appl. Math.}}
  {\bf{7}} 2 1954

\bibitem{nicolas} J.P. Nicolas, {\it{C. R. Acad. Sci. Paris, Ser.}} {\bf{1}} 344 2007.
  
  
\bibitem{sur} F. Colombini, E. De Giorgi, S. Spagnolo {\it{Ann. della Scuola Normale Superiore di Pisa }} {\bf{6}} 3 1979

\bibitem{lerner} F. Colombini, N. Lerner {\it{Duke Math. Journal}} {\bf{77}} 3 1995

\bibitem{men} F. Colombini, G. M\'etivier {\it{ArXiv:0611426}}

\bibitem{rud} C. Garetto, M. Ruzhansky {\it {Archive for Rational Mechanics and Analysis}} {\bf{217}} 1 2015




\bibitem{maeda}T. Shiromizu, K. Maeda, M. Sasaki,  {\it{Phys. Rev. D}} {\bf{ 62}}, 024012 2000






  
\bibitem{strin} A.Hebecker, J. March-Russell  {\it{Nucl. Phys. B}} {\bf{608}} 375–393 2001

\bibitem{bla1} J. Lehners, P. Madden, and N. Turok  {\it{Phys. Rev. D}} {\bf{75}}, 103510 2007

\bibitem{conical} J. P. Wilson  {\it{Class. Quantum Grav.}} {\bf{17}} 3199 2000



\end{thebibliography}
\end{document}